\documentclass[journal]{IEEEtran}
\usepackage{fancyhdr}
\usepackage{amsmath,amssymb}
\usepackage{amsfonts}
\usepackage[dvips]{graphicx}
\usepackage{dsfont}
\usepackage{balance}
\usepackage{algorithm}

\usepackage[hidelinks]{hyperref}    

\usepackage{color}
\usepackage{pgfplots}           
\pgfplotsset{compat=1.16}
\usepackage{enumerate}
\usepackage{graphics}
\usepackage{subfigure}
\usepackage{cite}
\usepackage{mathrsfs}
\usepackage{stfloats}

\usepackage{tikz}
\usepackage{mathdots}
\usepackage{yhmath}
\usepackage{cancel}
\usepackage{color}
\usepackage{siunitx}
\usepackage{array}
\usepackage{multirow}
\usepackage{textcomp}
\usepackage{gensymb}
\usepackage{tabularx}
\usepackage{booktabs}
\usepackage{graphicx}
\usepackage{xcolor}

\usepackage{enumitem}

\newtheorem{corollary}{Corollary}

\newtheorem{proposition}{Proposition}
\newtheorem{remark}{Remark}

\begin{document}

\vspace{-1.7cm}\title{Synergetic UAV-RIS Communication with Highly Directional Transmission}
\author{
Dimitrios Tyrovolas,~\IEEEmembership{Student Member,~IEEE,}
Sotiris A. Tegos,~\IEEEmembership{Student Member,~IEEE,}\\
Panagiotis D. Diamantoulakis,~\IEEEmembership{Senior Member,~IEEE,}
and George K. Karagiannidis,~\IEEEmembership{Fellow,~IEEE}
\thanks{The authors are with the Wireless Communications and Information Processing (WCIP) Group, Electrical \& Computer Engineering Dept., Aristotle University of Thessaloniki, 54 124, Thessaloniki, Greece (e-mails: \{tyrovolas,tegosoti,padiaman,geokarag\}@auth.gr).}
}
\maketitle
%

\begin{abstract} 
The effective integration of unmanned aerial vehicles (UAVs) in future wireless communication systems depends on the conscious use of their limited energy, which constrains their flight time. Reconfigurable intelligent surfaces (RISs) can be used in combination with UAVs with the aim to improve the communication performance without increasing complexity at the UAVs' side. In this paper, we propose a synergetic UAV-RIS communication system, utilizing a UAV with a highly directional antenna aiming to the RIS. The proposed scenario can be applied in all air-to-ground RIS-assisted networks and numerical results illustrate that it is superior from the cases where the UAV utilizes either an omnidirectional antenna or a highly directional antenna aiming towards the ground node.

\end{abstract}



\vspace{0.07in}
\section{Introduction}

	\IEEEPARstart{O}{ne} of the major requirements for future wireless communication networks is the effective integration of unmanned aerial vehicles (UAVs) \cite{bithas}. UAVs are envisioned to be used in a twofold way, i.e., either as
i) mobile equipment  for a vast number of operations including sensing and monitoring or ii) as part of the network's infrastructure for coverage extension, traffic offloading in crowded environments, and rapid recovery of the network services in cases of emergency. In both use cases, the effective utilization of UAVs depends on mainly two interrelated factors, the communication performance, e.g.,  in terms of data rate, reliability, and latency, as well as the UAVs' flight time duration which is limited by their battery capacity. Thus, it becomes of paramount importance to increase the  communication quality-of-service (QoS), without increasing the energy consumption at the UAVs. To this end, UAVs can be used in combination with reconfigurable intelligent surfaces (RISs), which can facilitate the signal beamforming through elements that can shift the phase of the reflected signals \cite{tegos}, without  increasing  complexity at the UAVs' side in order to meet the requirements of the aforementioned applications. \color{black} A potential synergetic UAV-RIS communication network could face the blockage problem efficiently by exploiting the characteristics of UAV channels as well as the passive beamforming gain achieved by RISs \cite{wuUAV}. Moreover, the utilization of RIS with UAVs has been examined as a possible way to achieve Ultra-Reliable Low Latency Communications (URLLC)\cite{Li,kaddoum}\color{black}. 

Considering that a RIS will be a stable target with non-negligible dimensions and high beamforming capabilities, UAVs can directively transmit towards the RIS in order to avoid applying complex signal processing algorithms for beamforming towards the GN which would lead to energy consumption increase and thus, restricting the UAVs' functionality. \color{black}The utilization of directional antennas in UAV-assisted communications has been proved to enhance coverage, therefore a synergetic UAV-RIS system is expected to provide better results \cite{gain}\color{black}. Most of the UAV-RIS works however, utilize UAVs equipped with omnidirectional antenna resulting in increased beam waste which could be overcome with a highly directional antenna steered towards the RIS\cite{direnzo}.  Existing link budget models though, do not adequately describe the losses in the scenario where a highly directional antenna is steered towards a RIS. Although the losses in RIS-aided communication systems in highly-directional millimeter wave links were partially investigated in \cite{ntontin},  the derived results may be proven inaccurate in UAV communication scenarios where the UAV can be arbitrarily located in the three-dimensional (3D) space. This is due to the fact that the proposed model in \cite{ntontin} is based on the assumptions that the radiation footprint and the RIS axes and center always coincide and either the radiation footprint or the RIS is a subarea of each other. Moreover, to fully investigate the potential of using UAVs and RISs in a synergetic way, except of the outage probability (OP), the temporal variations of the outage also need to be taken into account. These variations are characterized by the average outage duration (AOD), which is a second order statistics metric, also known as average fade duration. The AOD  is of paramount importance for all communication systems and especially for URLLC, because it facilitates the Markov modeling of wireless channels and determines the optimal packet length and packet error rate \cite{Alouini}. Furthermore, the authors in \cite{liu} model the RIS as a planar array, which cannot be utilized for the calculation of the aforementioned metrics, since the end-to-end channel is assumed to be free-space and the small-scale fading is absent.  Nevertheless, to the best of the authors' knowledge, the investigation of second order statistics has never been attempted in the context of RIS-assisted communication systems.

\color{black}To address the aforementioned issues, in this paper, a synergetic UAV-RIS communication system is proposed and its downlink performance is analyzed, utilizing a UAV with a directional antenna aiming to the RIS \color{black}. Specifically, we provide a link budget analysis, deriving the average received signal to noise ratio (SNR) as a function of the UAV's position in the 3D space as well as closed-form expressions for both OP and AOD. Furthermore, we provide numerical results which illustrate the impact of the distance on the average received SNR and the improved performance of the proposed synergetic network compared to the cases where either the UAV is equipped with an omnidirectional antenna or the higlhy directional antenna is steered towards the ground node (GN). Finally, the boundaries of the area where the UAV is permitted to hover into in order to ensure that the AOD is lower than a predefined threshold are illustrated.\vspace{0.15in}

\section{System Model}

We consider a communication network consisting of a UAV, a RIS and a GN equipped with a single omnidirectional antenna. 
The UAV is equipped with a highly-directional antenna steered towards the RIS with beamwidth less than or equal to $15^{\degree}$ \cite{ntontin}. 
The antenna's radiation pattern can be considered as a cone with fixed spreading angle and base lying upon the RIS. Due to the antenna's high directivity, no direct link between the UAV and the GN exists as the GN is not illuminated from the antenna's radiation pattern. This will stand, either if the UAV-GN link is blocked or not. The UAV is located at an arbitrary point $\left(r, \theta, \phi \right)$, where $r$ is the UAV-RIS center distance, $\theta \in \left[0 , \pi \right]$ is the elevation angle and $\phi \in \left[0 , \pi \right]$ is the azimuth angle. Moreover, the RIS is placed onto the plane $y=0$  and it is assumed that the UAV always aims to a proper point $K$ in order the radiation footprint's center to coincide with the RIS center $C$, which is considered as the origin of the axes, as illustrated in Fig. 1. The points $B$ and $D$ are the projection of the UAV on the plane $z=0$ and the projection of $B$ on the $x$ axis, respectively. In addition, $\phi'$ can be defined similarly to $\phi$ by replacing $C$ with $K$. According to the conic-section theory, the section of a cone and a rectangular plane is either a circle or an ellipse, thus the footprint upon the RIS plane is also a circle or an ellipse \cite{Hilbert}. Thus, the transmitted signal from the UAV antenna impinges upon the RIS which consists of $N$ reflecting elements and then it is reflected towards the GN. 

The RIS acts as a passive beamformer which adjusts the elements' reflection coefficient phase and shapes the transmitted signal in a desired way. However, as the transmission is highly directional, the number of the reflecting elements inside the radiation footprint, i.e., illuminated reflecting elements $M$, may be less than the total number of the reflecting elements $N$. Thus, the received signal at the GN, $Y$, can be expressed as
\begin{equation}
Y =  \sqrt{l_0 G P_t} \sum_{i=1}^{M} \lvert H_{i1} \rvert \lvert H_{i2} \rvert e^{-j \left( \omega_i + \arg(H_{i1}) + \arg(H_{i2}) \right)}X + W ,
\end{equation}
where $X$ is the transmitted signal for which it is assumed that $\mathbb{E}[|X|^2]=1$ with $\mathbb{E}[\cdot]$  and $\arg(\cdot)$ denoting expectation and the argument of a complex number, respectively. Also, $P_t$ denotes  the transmit power, $G = G_t G_r$ is the product of the UAV and the GN antenna gains,  and $H_{i1}$ and $H_{i2}$ are the complex channel coefficients that correspond to the $i$-th UAV-RIS and RIS-GN link, respectively.  Moreover, $W$ is the additive white Gaussian noise, $\omega_i$ is the phase correction term induced by the i-th reflecting element, and $l_0 = l_1 l_2$ with $l_1$ and $l_2$ being the path losses that correspond to the UAV-RIS and RIS-GN links, respectively. More specifically, $l_1$ equals to the fraction of the reflecting element's effective aperture to the radiation footprint area and $l_2$ can be modeled as  $l_2 = C_0 \left( \frac{d_u}{d_0} \right)^{-n}$, where $C_0$ denotes the reference path loss at the reference distance $d_0$, $d_u$ denotes the distance of the RIS-GN link and $n$ expresses the path loss exponent \cite{Wu}. Furthermore, it should be highlighted that due to the UAV's highly directional antenna and the UAV-RIS channel's nature which corresponds to an air to air channel, there is no fading in the UAV-RIS link, thus $\lvert H_{i1} \rvert = 1$ and $\arg(H_{i1})=\frac{2\pi r_i}{\lambda}$ with $\lambda$ being the carrier's frequency wavelength and $r_i$ the distance between the UAV and the $i$th-reflecting element. Additionally, it is assumed that $\lvert H_{i2} \rvert$ is a random variable (RV) following the Nakagami-$m$ distribution with shape parameter $m$ and spread parameter $\Omega$, which can describe accurately realistic communication scenarios characterized by severe or light fading. Also, each reflecting element adjusts the phase perfectly in order to cancel the overall phase shift, i.e., $\omega_i=- \frac{2\pi r_i}{\lambda} - \arg(H_{i2})$. Thus, the received signal at the GN can be rewritten as
\begin{equation}
	Y =  \sqrt{l_0 G P_t} H X + W ,
\end{equation}
where $H =  \sum_{i=1}^{M} \lvert H_{i2} \rvert$.
\color{black} By utilizing \cite{justin}, the system model can be extended and describe the case where the phase correction term is not chosen perfectly thus, the overall phase shift is not nullified. \color{black} 

\begin{figure}
\centering
\includegraphics[width=0.55\linewidth]{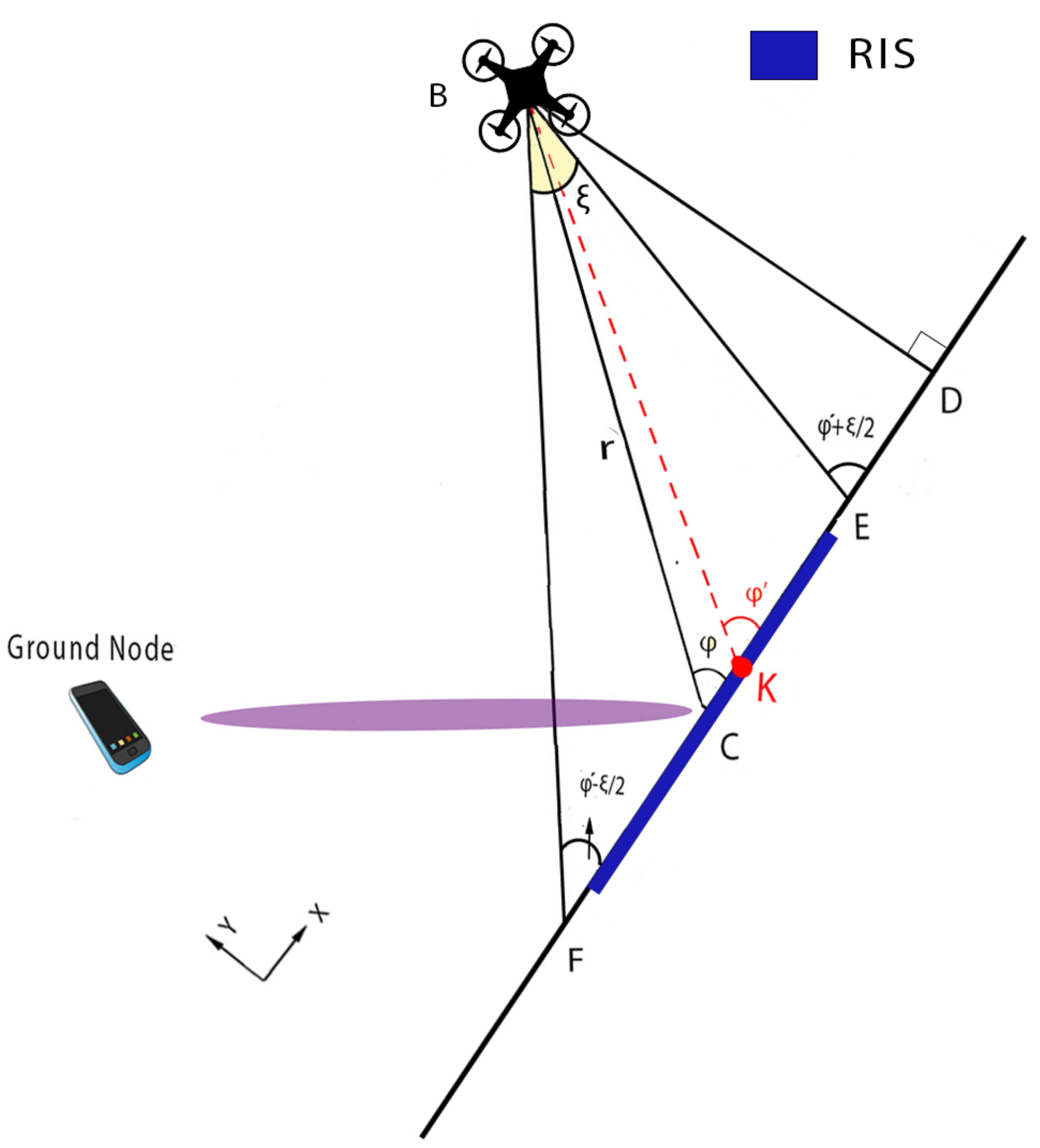}
  \caption{The synergetic UAV-RIS communication system's layout.}
  \label{fig:Geometrical}
\end{figure}

\section{Performance Analysis}
In this section, we present the average received SNR, the OP and the AOD for the proposed synergetic UAV-RIS system. 

\subsection{Link Budget Analysis}
The instantaneous received SNR at the GN is given by
\begin{equation} \label{inst_SNR}
\gamma_r = l_0 G \gamma_t \lvert H \rvert ^{2},
\end{equation}
where $\gamma_t = \frac{P_t}{\sigma ^{2}}$ is the transmit SNR and $\sigma^{2}$ denotes the noise power. In \eqref{inst_SNR}, we need to determine the path loss $l_0$ of the proposed synergetic system and to investigate the channel gain through the calculation of the number of the illuminated elements. To this end, we first define
\begin{equation}
\begin{split}
g(x) &= \Bigg \lvert \frac{r \sin(\phi) \sin(\theta)}{\tan\left(\phi\right)} - \frac{r \sin(\phi) \sin(\theta)}{\tan\left(x + \frac{\xi}{2}\right)} \Bigg \rvert \\
& - \Bigg \lvert \frac{r \sin(\phi) \sin(\theta)}{\tan\left(x - \frac{\xi}{2}\right)} - \frac{r \sin(\phi) \sin(\theta)}{\tan\left(\phi\right)} \Bigg \rvert,
\end{split}
\end{equation}
where $x\in [ \phi, \frac{\pi}{2} ] $ if $\phi \leq \frac{\pi}{2}$ and $x\in ( \frac{\pi}{2}, \phi ] $ if $\phi > \frac{\pi}{2}$.

In the following proposition, we provide the path loss of the link between the UAV and the RIS.
\begin{proposition}
The path loss of the UAV-RIS link can be expressed as
\begin{equation}
l_1 = \frac{S}{\pi r^{2}}  \left( \Bigg \lvert \frac{ \sin(\phi) \sin(\theta)}{\tan\left(\phi\right)} - \frac{ \sin(\phi) \sin(\theta)}{\tan\left(\phi' + \frac{\xi}{2}\right)} \Bigg \rvert  \tan\left(\frac{\xi}{2} \right) \right)^{-1},
\end{equation}
where $S$ denotes the effective aperture of the reflecting element which equals to its area and $\phi'$ can be calculated with numerical methods as the root of $g(\phi') = 0$ . 
\end{proposition}
\begin{IEEEproof}
See Appendix A.
\end{IEEEproof}

\begin{figure}
  \centering
  \includegraphics[width=0.55\linewidth]{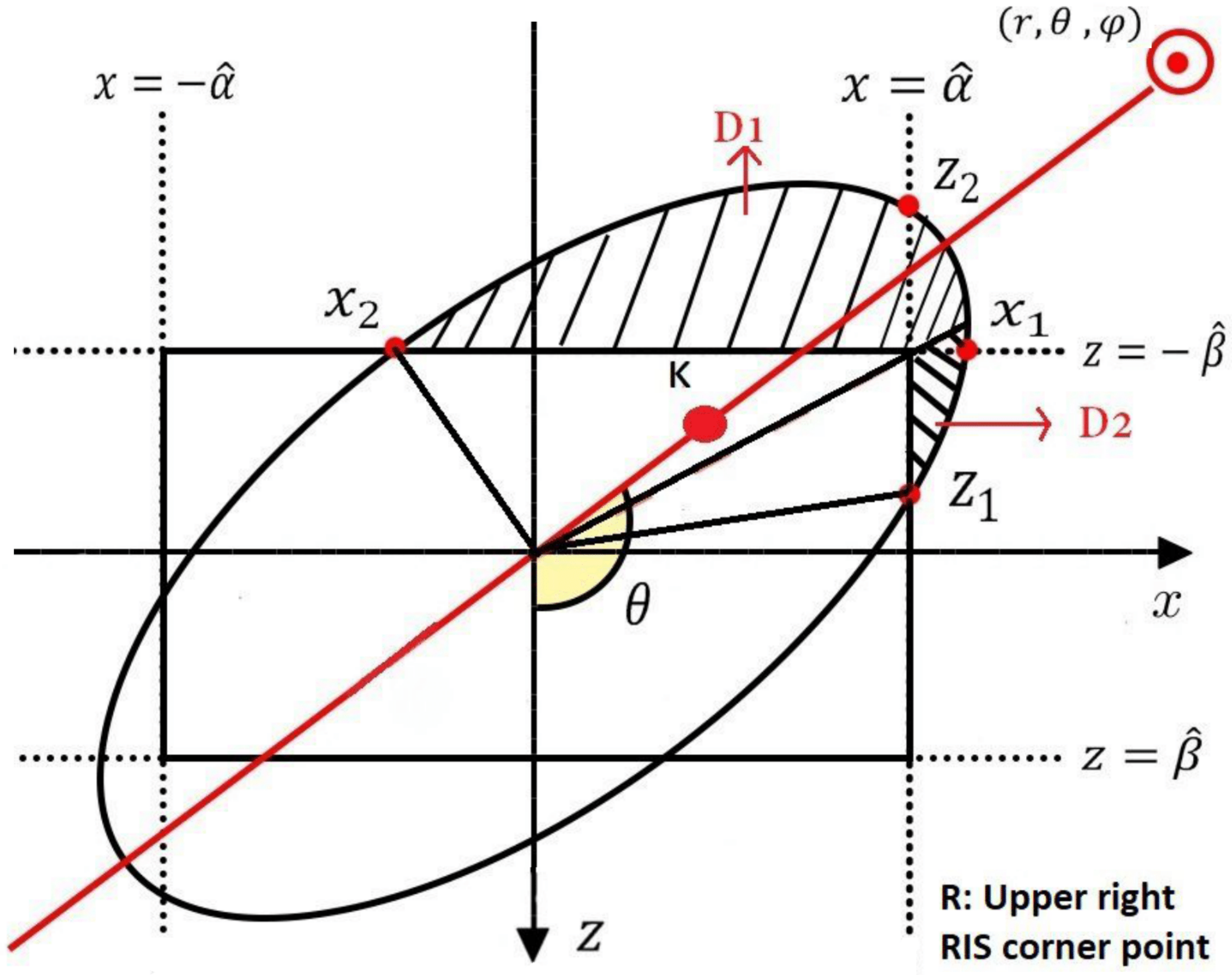}
  \caption{Radiation footprint  at plane $y=0$ and UAV's location at plane $y = r \sin(\theta) \sin(\phi)$.}
  \label{fig:Footprint}
\end{figure}

After the calculation of the footprint area, considering the position of the UAV, there can be five cases regarding the radiation footprint and its intersection points with the RIS sides. It should be highlighted that only the areas $D_1$ and $D_2$, which are illustrated in Fig. 2, need to be calculated due to the geometrical problem's symmetry. Considering that $x_1$, $x_2$ and $z_1$, $z_2$ are the intersection point pairs of the footprint with the lines that define the  upper and the right RIS sides, respectively, as shown in Fig. 2, the cases are the following:
\begin{itemize}
\item ${C}_1$: No intersection points exist, i.e., the footprint is inside the RIS, thus no spillover losses exist.
\item ${C}_2$: Intersection points $k_1, k_2$  exist only in one RIS side where $k \in \{x,z\}$.
\item ${C}_3$: All intersection points exist and are on the RIS sides.
\item ${C}_4$: All intersection points exist and only one of each pair lies on the RIS side.
\item ${C}_5$: All intersection points exist and are beyond the RIS sides, i.e., the RIS is inside the footprint, thus all the reflecting elements are being illuminated.
\end{itemize}

By taking into account the aforementioned cases, in the following proposition we provide the number of illuminated reflecting elements as well as the spillover losses which correspond to the radiation footprint parts that lay outside of the RIS to calculate the average channel gain of the proposed synergetic system. To this end, we first define $E_{QV} =\frac{(t_{2} - t_{1}) \alpha(\xi) \beta(\xi)}{2}$ which expresses the area of an elliptic arc which is defined by the origin and two points $Q\equiv (m_1,n_1)$ and $V\equiv(m_2,n_2)$, where $V$  is located counter-clockwise related to $Q$ \cite{hughes}. Moreover, $t_{1}$ is the angle that is shaped with the major axis and the line that connects the origin with $Q$ and $t_{2}$ is the angle that is shaped similarly with $V$. These angles can be calculated as in Table $1$ in \cite{hughes}. Finally,  $T_{QV} = \frac{ \lvert m_1 n_2 -m_2 n_1 \rvert}{2} $ denotes a triangle area whose vertex coincides with the origin and base is formed from the points $Q\equiv (m_1,n_1)$ and $V\equiv(m_2,n_2)$  \cite{hughes}.
\begin{proposition}
The channel coefficient $ H$ can be approximated with $\tilde H$ which follows the normal distribution with parameters $\mathbb{E}[ \tilde H ] = M\frac{\Gamma(m+\frac{1}{2})}{\Gamma(m)} \left(\frac{\Omega}{m}\right)^\frac{1}{2}$ and $ \mathrm{Var}[ \tilde H ] = M \Omega \left(1-\frac{1}{m}\left(\frac{\Gamma(m+\frac{1}{2})}{\Gamma(m)}\right)^2\right)$, where  $\Gamma(\cdot)$ is the gamma function and $M  = \left\lfloor \frac{E_f  \left(1-J\right)}{4 \hat{\alpha} \hat{\beta}} N \right\rfloor$ with
\begin{equation}   
J = 
     \begin{cases}
     0, &\text{if ${C}_1$}\\
     \frac{2 \left( E_{k_1 k_2} - T_{k_1 k_2} \right)}{E_f}, &\text{if ${C}_2$}\\
     \frac{2 \left(E_{x_1 x_2} - T_{x_1 x_2} + \mathrm{E}_{z_1 z_2} - T_{z_1 z_2}\right) }{E_f}, &\text{if ${C}_3$}\\
     \frac{2\left(E_{z_1 x_2} - T_{R x_2}  - T_{z_1 R}\right) }{E_f}, &\text{if ${C}_4$}\\
     1 - \frac{4 \hat{\alpha} \hat{\beta} }{E_f}, &\text{if ${C}_5$}\\
     \end{cases} 
\end{equation}
denoting the percentage of the footprint area which is lost due to spillover losses and $\lfloor \cdot \rfloor$ being the floor operator.
\end{proposition}

\begin{IEEEproof}
See Appendix B.
\end{IEEEproof}

\begin{remark}
Considering that $\tilde H$ follows the normal distribution,  $\left | \frac{\tilde H}{\sqrt{ \mathrm{Var}[ \tilde H ]}}\right |  ^2$ is a RV following the non-central chi-square distribution with one degree of freedom and non centrality parameter equal to $\frac{\mathbb{E}^{2}[\tilde H]}{ \mathrm{Var}[ \tilde H ]}$, thus the average received SNR can be calculated as 
\begin{equation}
\mathbb{E}[\gamma_r] = l_0 G \gamma_t \left( \mathbb{E}^{2}[\tilde H] + \mathrm{Var}[ \tilde H ]\right).
\end{equation}
\end{remark}

\subsection{Outage Probability and Average Outage Duration}
Next, we derive the OP, which is defined as the probability that the instantaneous received SNR is below a specified threshold, in order to evaluate the considered system's reliability. To this end, we set the parameter $z =  \sqrt{\frac{\gamma_{\mathrm{thr}}}{l_0 G \gamma_t }} $, where $\gamma_{\mathrm{thr}}$ is the outage threshold value of the received SNR.
\begin{corollary} \label{P_out}
The OP can be obtained as
\begin{equation}
\begin{split}
\mathcal{P}_o (z)=  \frac{1}{2} & \left[1 + \mathrm{erf}\left(\frac{z -\mathbb{E}[ \tilde H ] }{\sqrt{2\mathrm{Var}[ \tilde H ]}}\right)  \right], 
\end{split}
\end{equation}
\end{corollary}
where $\mathrm{erf}(\cdot)$ is the error function,
\begin{IEEEproof}
The OP is defined as
\begin{flalign}
\mathcal{P}_o(z) = \Pr \left(\gamma_r \leq \gamma_{\mathrm{thr}} \right) = \Pr \left(  H \leq z \right).
\end{flalign}
Considering the approximation $\tilde H$ which follows the normal distribution, we can calculate the OP through the cumulative density function and derive (8) which completes the proof.
\end{IEEEproof}

In the following corollary, we provide the AOD which is defined as the average time that the fading envelope remains below a specified level after crossing that level in a downward direction, characterizing the temporal variation of the outage probability and is closely related to the system’s delay. The AOD is calculated as the fraction of the outage probability at a threshold $z$ to the level crossing rate (LCR) at $z$ which is defined as the rate at which the received signal crosses the threshold $z$ in the negative direction.
\begin{corollary}
The AOD can be calculated as
\begin{equation} \label{AOD}
A \left( z \right) = \frac{2 \mathcal{P}_o(z) \sqrt{m \mathrm{Var}[ \tilde H ]}}{ e^{ -\frac{\left(z-\mathbb{E}[ \tilde H ]\right)^2}{2 \mathrm{Var}[ \tilde H ]}} f_{D} \sqrt{M \Omega}},
\end{equation}
 where $f_{D}$ is the maximum Doppler spread.
\end{corollary}
\begin{IEEEproof}
See Appendix C.
\end{IEEEproof} 
\vspace{-0.1in}
\section{Simulation Results}
\color{black}In this section, we illustrate the performance of the proposed synergetic UAV-RIS system and how it is affected by the UAV-RIS distance\color{black}. We examine two communication scenarios which correspond to a LoS and a non-LoS (nLoS) scenario between the RIS and the GN, where we set the shape parameter of the second link $m=3$ for the first scenario and $m=1$ for the second one and the spreading parameter $\Omega=1$ for both cases. For the second link, we set $C_0= -30$ dB, $d_0= 1$ m, $d_u = 50$ m and the path loss exponent $n= 2.2$ for both scenarios.  \color{black}Furthermore, we set the transmit SNR $\gamma_t=  50$ dB, the threshold value $\gamma_{\mathrm{thr}}=20$ dB, the UAV's highly directional antenna spreading angle $\xi = 15^\degree$ as well as the azimuth angle $\phi = \frac{\pi}{2}$, unless it is stated otherwise. Moreover, we set the RIS dimensions as $2\hat{\alpha}= 2$ m and $2\hat{\beta}= 1$ m and the number of the reflecting elements $N = 800$ which are assumed to have square area with sub-wavelength dimensions\color{black}. It should be mentioned that the inter-distance between the elements is assumed to be $0$ and the mutual coupling between the reflecting elements is neglected. \color{black}Finally, we calculate the UAV's highly directional antenna gain as $G_t = \frac{29000}{\xi ^2}$ \cite{gain}, and considering that the GN antenna is omnidirectional, we set $G_r=1$. \color{black}

Fig. 3a illustrates the impact of the UAV-RIS distance to the average received SNR. As the distance increases, the average received SNR for the proposed scenario starts to increase due to the fact that more reflecting elements are being illuminated. However, after a certain distance the average received SNR increase slower due to spillover losses. The reason why the average received SNR continues to increase is due to the fact that the illuminated elements keep increasing as the UAV moves away from the RIS. When all the elements are illuminated, the spillover losses continue to increase as the distance $r$ increases leading to rapid decrease of the average received SNR. Moreover, to evaluate the performance of the proposed synergetic network with directional transmission, the average received SNR is compared with two benchmark scenarios where in the first one the UAV is equipped with an omnidirectional antenna, i.e., $G_t =1$, and the GN is served from both RIS and directly from the UAV, and in the second one the highly directional antenna is steered towards the GN and no RIS is utilized. It should be highlighted that the UAV-GN link for both benchmark scenarios is assumed to not be affected by small-scale fading which is considered as the best case scenario. Furthermore, the GN is located at the $x=0$ plane as the RIS center and the UAV and forms an angle of $45^{\degree}$ with the RIS center. Moreover, the path loss for UAV-RIS and UAV-GN links in the first benchmark is expressed as $l_2$ with $C_0 = \frac{A_e}{4 \pi}$ and $n=2$ where $A_e \in \{S, \frac{\lambda^2}{4 \pi} \}$ for the UAV-RIS and the UAV-GN link, respectively, while for the second benchmark the path loss is given as $\frac{\lambda^2}{4 \pi E_f'}$, where $E_f'$ is the radiation footprint that lays on the ground. The maximal average receiver SNR value of the proposed scenario is greater than the one of the benchmark scenarios, even if the RIS-GN link is characterized by Rayleigh fading, i.e., $m=1$.  

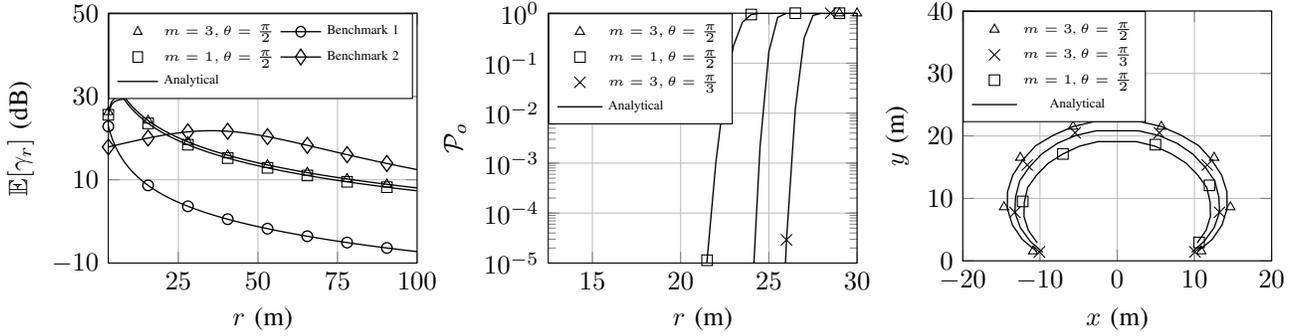
\begin{figure*}[t]
\centering
\begin{minipage}[t]{.32\textwidth}
	\centering
	\begin{tikzpicture}
	\begin{axis}[
	width=0.98\linewidth,
	xlabel = {$r$ (m)},
	ylabel = {$\mathbb{E}[\gamma_r]$ (dB)},
	xmin = 3,
	xmax = 100,
	ymin = -10,
	ymax = 50,
	ytick = {-10,10,...,50},
	xtick = {0,25,...,100},
	grid = major,
	legend image post style={xscale=0.9},
	legend columns=2, 
	legend cell align = {left},
      legend style={at={(0,1)},anchor=north west, font = \tiny},
	]
	\addplot[
	black,
          mark=triangle,
	mark repeat = 25,
	mark size = 2,
          only marks,
	]
	table {data1/apotelesmaM_pi2_3.dat};
	\addlegendentry{$m =3, \theta= \frac{\pi}{2}$}

	
	\addplot[
	black,
       mark = o ,
      mark repeat = 25,
      mark size =2,
	line width = 0.5pt,
	style = solid,
	]
	table {data1/apotelesmaMd_pi2_3.dat};
      \addlegendentry{Benchmark 1}
      
      		\addplot[
      black,
      mark=square,
      mark repeat = 25,
      mark size = 2,
      only marks,
      ]
      table {data1/apotelesmaM_pi2_1.dat};
      \addlegendentry{$m =1, \theta= \frac{\pi}{2}$}
      
	\addplot[
	black,
       mark = diamond ,
      mark repeat = 25,
      mark size =3,
	line width = 0.5pt,
	style = solid,
	]
	table {data1/apotelesmaMuser.dat};
      \addlegendentry{Benchmark 2}
	
	\addplot[
	black,
	no marks,
	line width = 0.5pt,
	style = solid,
	]
	table {data1/apotelesmaM_pi2_3.dat};
	\addlegendentry{Analytical}
	
	\addplot[
	black,
           no marks,
	line width = 0.5pt,
	style = solid,
	]
	table {data1/apotelesmaM_pi2_1.dat};
	\end{axis}
	\end{tikzpicture}
	\label{fig:distance}
\end{minipage}%
\begin{minipage}[t]{.32\textwidth}
	\centering
	\begin{tikzpicture}
	\begin{semilogyaxis}[
	width=0.98\linewidth,
	xlabel = {$r$ (m)},
	ylabel = {$\mathcal{P}_o$},
	xmin = 12.5,
	xmax = 30,
	ymin = 0.00001,
	ymax = 1,
	ytick = {0.00001,0.0001,0.001,0.01,0.1,1},
	xtick = {10,15,...,30},
	grid = major,
	legend cell align = {left},
      legend style={at={(0,1)},anchor=north west, font = \tiny}
	]
	\addplot[
	black,
    mark=triangle,
	mark repeat = 2.5,
	mark size = 2,
	mark phase = 10,
only marks,
	]
	table {data1/outage1_pi2_3.dat};
	\addlegendentry{$m =3, \theta= \frac{\pi}{2}$}
	\addplot[
	black,
    mark=square,
	mark repeat = 5,
	mark size = 2,
only marks,
	]
	table {data1/outage1_pi2_1.dat};
	\addlegendentry{$m =1, \theta= \frac{\pi}{2}$}
	\addplot[
	black,
    mark=x,
	mark repeat = 5,
	mark size = 3,
only marks,
	]
	table {data1/outage1_pi6_3.dat};
	\addlegendentry{$m =3, \theta= \frac{\pi}{3}$}

	\addplot[
	black,
          no marks,
          line width = 0.5pt,
	style = solid,
	]
	table {data1/outage1_pi2_3.dat};
	\addplot[
	black,
          no marks,
	line width = 0.5pt,
	style = solid,
	]
	table {data1/outage1_pi2_1.dat};
	\addplot[
	black,
          no marks,
	line width = 0.5pt,
	style = solid,
	]
	table {data1/outage1_pi6_3.dat};
	\addlegendentry{Analytical}
	\end{semilogyaxis}
	\end{tikzpicture}
	\label{fig:outage}
\end{minipage}%
\begin{minipage}[t]{.32\textwidth}
	\centering
	\begin{tikzpicture}
		\begin{axis}[
			width=0.98\linewidth,
			xlabel = $x$ (m),
			ylabel = $y$ (m),
			xmin = -20,xmax = 20,
			ymin = 0,
			ymax = 40,
			grid = major,
			legend style={at={(0,1)},anchor=north west, font = \tiny}
			]
			\addplot[
			black,
			mark=triangle,
			mark repeat = 3,
			mark size = 2,
			only marks,
			]
			table {data1/AFD_pi2s_3_1.dat};
			\addlegendentry{$m=3$, $\theta = \frac{\pi}{2}$}
			\addplot[
			black,
			mark=x,
			mark repeat = 3,
			mark size = 3,
			only marks,
			]
			table {data1/AFD_pi6s_3_1.dat};		    
			\addlegendentry{$m =3$, $\theta = \frac{\pi}{3}$}
			\addplot[
			black,
			mark=square,
			mark repeat = 4,
			mark size = 2,
			only marks,
			]
			table {data1/AFD_pi2s_1_1.dat};
			\addlegendentry{$m=1$, $\theta = \frac{\pi}{2}$}
			\addplot[
			black,
			no marks,
			line width = 0.5pt,
			style = solid,
			]
			table {data1/AFD_pi2_3_1.dat};
			\addplot[
			black,
			no marks,
			line width = 0.5pt,
			style = solid,
			]
			table {data1/AFD_pi6_3_1.dat};		    
			\addplot[
			black,
			no marks,
			line width = 0.5pt,
			style = solid,
			]
			table {data1/AFD_pi2_1_1.dat};
			\addlegendentry{Analytical}
		\end{axis}
	\end{tikzpicture}	
	\label{fig:areas}
\end{minipage}
\vspace{-0.2in}
\caption{a) Average received SNR versus UAV-RIS distance $r$. b) OP versus UAV-RIS distance $r$. c) Hovering permitted area boundaries for $\mathrm{AOD} = 1$ ms.}
\vspace{-0.2in}
\end{figure*}

Fig. 3b depicts how the OP is affected from the UAV-RIS distance. It can be observed that if $m = 3$ and $\theta =\frac{\pi}{2}$, i.e., the UAV is at the level of the RIS center, the UAV can ensure low OP for distances up to $20$ m. Thus, Fig. 3b demonstrates that the cooperation of UAVs equipped with highly directional antennas and RIS can potentially provide ultra reliable communications.

In Fig. 3c, we illustrate the area boundaries where the UAV is permitted to hover into, in order to ensure that the AOD, is less than $1$ ms. Due to the randomness of the wireless environment, the UAV can select a proper point in the hovering permitted area in order to maintain the AOD below $1$ ms as well as the LoS link. \color{black}The considered scenario is characterized by low-mobility and, thus, a relatively small value of maximum Doppler spread, i.e., $f_D=5$ Hz, has been selected. \color{black} As $\theta$ diverges from $\frac{\pi}{2}$ or the required AOD value decreases, the area that the UAV can hover into also decreases. Moreover the fading conditions of the RIS-GN link can also affect the AOD. Thus, in order to keep the AOD value low, the RIS should be placed properly in order to ensure good fading conditions between the RIS and the GN.

\vspace{-0.1in}
\section{Conclusions}
In this work, we have proposed a synergetic UAV-RIS system where a UAV equipped with a highly directional antenna acts as an aerial communication node that serves a GN with the aid of a RIS. Specifically, we have provided a link budget analysis deriving the average received SNR as a function of the UAV's position in the 3D space and numerical results which illustrate the superiority of the proposed system compared to two benchmark scenarios. Moreover, we have derived closed-form expressions for both the OP and the AOD which can be used to evaluate the system's reliability and latency. \color{black}The derived performance metrics can also describe the performance of the proposed synergetic UAV-RIS communication system in the multi-user case, considering an orthogonal multiple access scheme, e.g., Time Division Multiple Access (TDMA).\color{black}
\vspace{-0.2in}
\begin{appendices}
\section{Proof of Proposition 1}
Let the UAV be in an arbitrary location $\left( r, \theta, \phi \right)$ in the 3D space. According to Friis' formula for the received power in the far-field free space case, the received power is equal to the radiation pattern section that impinges upon the effective aperture of the receiver's antenna. Thus, the path loss of the UAV-RIS link $l_1 = \frac{S}{E_f}$ where $E_{f} = \pi \alpha \beta$  is the radiation footprint area with radii $\alpha$ and $\beta$. The ellipse's major radius can be calculated by utilizing the law of cosines as $  \alpha= (CD) - (ED) =\Bigg \lvert \frac{r \sin(\phi) \sin(\theta)}{\tan\left(\phi\right)} - \frac{r \sin(\phi) \sin(\theta)}{\tan\left(\phi' + \frac{\xi}{2}\right)} \Bigg \rvert .$ The angle $\phi'$ can be calculated numerically by equalizing the line segments $(CE)$ and $(FC)$ as shown in Fig. 1 and considering that $g(\phi')$ = $(CE) - (FC)$. As shown in Fig. 2, the points of the major axis have elevation angle $\theta$ and azimuth angle $0$ or $\pi$ since they are located at the plane $y=0$. Furthermore, the first coordinate of the aiming point  is calculated as $(CD)$ - $(KD)$ as shown in Fig. 1, thus $K\equiv \left( \lvert r \sin\left(\theta\right) \cos\left(\phi\right) - \frac{r\sin(\theta)\sin(\phi)}{\tan(\phi')} \rvert, \theta, \varsigma \right)$ with  $\varsigma =0$ if $\phi \leq\frac{\pi}{2}$ or $\varsigma =\pi$ if $\phi >\frac{\pi}{2}$. Moreover, the minor axis coincides with the base of the isosceles triangle with oppose angle $\xi$ and height $r$, thus the minor radius can be expressed as $\beta = r \tan\left(\frac{\xi}{2}\right)$. After the calculation of the radii, we can calculate the radiation footprint area $E_{f}$ and the path loss of the UAV-RIS link $l_1$ which concludes the proof.
\vspace{-0.2in}
\section{Proof of Proposition 2}
Due to the directional transmission, the incident power at the RIS is equal to the percentage of the transmit power that lays inside the RIS. Utilizing the lines that define the RIS sides and the equation of a rotated ellipse which is given by
$\frac{ \left( x \sin(\theta) + y \cos(\theta) \right)^{2}}{\alpha^{2}(\xi)} + \frac{ \left( x \sin(\theta) - y \cos(\theta) \right)^{2}}{\beta^{2}(\xi)} = 1 $,
we can determine the intersection points $x_1$, $x_2$, $z_1$ and $z_2$. Considering  Fig. \ref{fig:Footprint} and the different cases regarding the intersection points of the radiation footprint with the RIS sides,the corresponding  arc areas and  triangles are calculated.
Next, the area of the footprint that lies outside of the RIS is calculated which enables the derivation of $M$. Since the UAV always hovers in a location where a large number of reflecting elements is illuminated, i.e., $M \gg 1$, according to the central limit theorem, $H$ converges to a Gaussian distributed random variable $\tilde H$.
\vspace{-0.2in}
\section{Proof of Corollary 2}
The AOD is defined as \cite{Yacoub} $A \left( z \right) = \frac{\mathcal{P}_o(z)}{L(z)}$. For an arbitrary stationary differentiable random process $x(t)$, the LCR is given by
$L(z) = \int_{-\infty}^{\infty} \lvert \dot{x} \rvert f_{X,\dot{X}}(z, \dot{x}) d\dot{x}$ ,
where $\dot{x}$ is the first derivative of $x$ with respect to time and $f_{X,\dot{X}}(z, \dot{x})$ is the joint probability density function (PDF) of $x(t)$ and $\dot{x}(t)$ at time $t$  \cite{Yacoub}.
Considering that $H$ is a sum of $M$  RVs following the Nakagami-$m$ distribution, the LCR can be expressed as $ L(z) = f_{H}(z) \frac{\sigma_{\dot{x}}}{\sqrt{2 \pi}}$, where $\sigma_{\dot{x}} = \pi f_{D} \sqrt{\frac{M \Omega}{m}}$ and $f_H(z)$ denotes the PDF of  $H$ \cite{mathiop}. Approximating $f_{H}(z) $  with the PDF of  $\tilde{H}$ which is given by $f_{\tilde H}(z) = \frac{1}{\sqrt{2\pi  \mathrm{Var}[ \tilde H ]}} e^{ -\frac{\left(z-\mathbb{E}[ \tilde H ]\right)^2}{2 \mathrm{Var}[ \tilde H ]}}$,   
\eqref{AOD} is derived, which completes the proof.
\end{appendices}
\vspace{-0.1in}
\bibliographystyle{IEEEtran}
\bibliography{Bibliography}

\begin{thebibliography}{10}
\providecommand{\url}[1]{#1}
\csname url@samestyle\endcsname
\providecommand{\newblock}{\relax}
\providecommand{\bibinfo}[2]{#2}
\providecommand{\BIBentrySTDinterwordspacing}{\spaceskip=0pt\relax}
\providecommand{\BIBentryALTinterwordstretchfactor}{4}
\providecommand{\BIBentryALTinterwordspacing}{\spaceskip=\fontdimen2\font plus
\BIBentryALTinterwordstretchfactor\fontdimen3\font minus
  \fontdimen4\font\relax}
\providecommand{\BIBforeignlanguage}[2]{{%
\expandafter\ifx\csname l@#1\endcsname\relax
\typeout{** WARNING: IEEEtran.bst: No hyphenation pattern has been}%
\typeout{** loaded for the language `#1'. Using the pattern for}%
\typeout{** the default language instead.}%
\else
\language=\csname l@#1\endcsname
\fi
#2}}
\providecommand{\BIBdecl}{\relax}
\BIBdecl

\bibitem{bithas}
P.~S. Bithas, V.~Nikolaidis, A.~G. Kanatas, and G.~K. Karagiannidis,
  ``{UAV-to-Ground Communications: Channel Modeling and UAV Selection},''
  \emph{IEEE Trans. Wireless Commun.}, vol.~68, no.~8, pp. 5135--5144, Aug.
  2020.

\bibitem{tegos}
\BIBentryALTinterwordspacing
S.~A. Tegos, D.~Tyrovolas, P.~D. Diamantoulakis, and G.~K. Karagiannidis, ``{On
  the Distribution of the Sum of Double-Nakagami-m Random Vectors and
  Application in Randomly Reconfigurable Surfaces},'' 2021. [Online].
  Available: \url{https://arxiv.org/abs/2102.05591}
\BIBentrySTDinterwordspacing

\bibitem{wuUAV}
\BIBentryALTinterwordspacing
Q.~Wu, J.~Xu, Y.~Zeng, D.~W.~K. Ng, N.~Al-Dhahir, R.~Schober, and A.~L.
  Swindlehurst, ``{A Comprehensive Overview on 5G-and-Beyond Networks with
  UAVs: From Communications to Sensing and Intelligence},'' 2021. [Online].
  Available: \url{arxiv.org/abs/2010.09317}
\BIBentrySTDinterwordspacing

\bibitem{Li}
Y.~Li, C.~Yin, T.~Do-Duy, A.~Masaracchia, and T.~Q. Duong, ``{Aerial
  Reconfigurable Intelligent Surface-Enabled URLLC UAV Systems},'' \emph{IEEE
  Access}, vol.~9, pp. 140\,248--140\,257, 2021.

\bibitem{kaddoum}
{Ranjha, Ali and Kaddoum, Georges}, ``{URLLC Facilitated by Mobile UAV Relay
  and RIS: A Joint Design of Passive Beamforming, Blocklength, and UAV
  Positioning},'' \emph{IEEE Internet Things J.}, vol.~8, no.~6, pp.
  4618--4627, 2021.

\bibitem{gain}
H.~Shakhatreh, A.~Khreishah, N.~S. Othman, and A.~Sawalmeh, ``{Maximizing
  indoor wireless coverage using UAVs equipped with directional antennas},'' in
  \emph{2017 IEEE 13th MICC}, Nov. 2017, pp. 175--180.

\bibitem{direnzo}
S.~Li, B.~Duo, X.~Yuan, Y.-C. Liang, and M.~Di~Renzo, ``{Reconfigurable
  Intelligent Surface Assisted UAV Communication: Joint Trajectory Design and
  Passive Beamforming},'' \emph{IEEE Wireless Commun. Lett.}, vol.~9, no.~5,
  pp. 716--720, 2020.

\bibitem{ntontin}
K.~Ntontin, A.-A.~A. Boulogeorgos, D.~G. Selimis, F.~I. Lazarakis, A.~Alexiou,
  and S.~Chatzinotas, ``{Reconfigurable Intelligent Surface Optimal Placement
  in Millimeter-Wave Networks},'' \emph{IEEE OJ-COMS}, vol.~2, pp. 704--718,
  2021.

\bibitem{Alouini}
C.~B. Issaid and M.-S. Alouini, ``{Level Crossing Rate and Average Outage
  Duration of Free Space Optical Links},'' \emph{IEEE Trans. Commun}, vol.~67,
  no.~9, pp. 6234--6242, Sep. 2019.

\bibitem{liu}
\BIBentryALTinterwordspacing
Y.~Liu, B.~Duo, Q.~Wu, X.~Yuan, J.~Li, and Y.~Li, ``{Elevation Angle-Dependent
  Trajectory Design for Aerial RIS-aided Communication},'' 2021. [Online].
  Available: \url{arxiv.org/abs/2108.10180#}
\BIBentrySTDinterwordspacing

\bibitem{Hilbert}
D.~Hilbert and S.~Cohn-Vossen, ``{Geometry and the imagination},''
  \emph{NewYork: Chelsea Publishing Company}, 1952.

\bibitem{Wu}
Q.~Wu and R.~Zhang, ``{Intelligent Reflecting Surface Enhanced Wireless Network
  via Joint Active and Passive Beamforming},'' \emph{IEEE Trans. Wireless
  Commun}, vol.~18, no.~11, pp. 5394--5409, Nov. 2019.

\bibitem{justin}
M.-A. Badiu and J.~P. Coon, ``{Communication Through a Large Reflecting Surface
  With Phase Errors},'' \emph{IEEE Wireless Commun. Lett.}, vol.~9, no.~2, pp.
  184--188, 2020.

\bibitem{hughes}
G.~B. Hughes and M.~Chraibi, ``{Calculating ellipse overlap areas},''
  \emph{Comput. Visual. Sci.}, vol.~15, no.~5, pp. 291--301, Oct. 2012.

\bibitem{Yacoub}
M.~D. {Yacoub}, J.~E.~V. {Bautista}, and L.~{Guerra de Rezende Guedes}, ``{On
  higher order statistics of the Nakagami-m distribution},'' \emph{IEEE Trans.
  Veh. Technol.}, vol.~48, no.~3, pp. 790--794, May 1999.

\bibitem{mathiop}
C.-D. Iskander and P.~Takis~Mathiopoulos, ``{Analytical level crossing rates
  and average fade durations for diversity techniques in Nakagami fading
  channels},'' \emph{IEEE Trans. Commun.}, vol.~50, no.~8, pp. 1301--1309, Aug.
  2002.

\end{thebibliography}


\end{document}